\documentclass[12pt]{article}

\usepackage{amsmath,amssymb,empheq,hyperref}

\newcommand{\ie}{{\it i.e.\/}, }
\newcommand{\eg}{{\it e.g.\/}, }

\newcommand{\set}[1]{\mathrm{#1}}

\newcommand{\RR}{\mathbb{R}}
\newcommand{\CC}{\mathbb{C}}
\newcommand{\iu}{\mathrm{i}}

\newcommand{\lexp}[1]{\mathrm{e}^{#1}}

\newcommand{\RE}{\mathop{\set{Re}}}
\newcommand{\diff}{\mathop{\mathrm{\mathstrut{d}}}\!}
\newcommand{\diag}{\mathop{\mathrm{diag}} }

\newcommand{\Sob}[1]{\set{H}^{#1}}
\newcommand{\C}[1]{\set{C}^{#1}}

\newcommand{\A}{\mathcal{A}}
\newcommand{\B}{\mathcal{B}}
\newcommand{\T}{\mathcal{T}}
\newcommand{\R}{\mathcal{R}}
\newcommand{\N}{\mathcal{N}}
\newcommand{\LL}{\mathcal{L}}
\newcommand{\lu}{\ell}
\newcommand{\cS}{\mathcal{G}}
\newcommand{\cs}{g}
\newcommand{\cSpm}{\mathcal{G}^{\pm}}
\newcommand{\Y}{\mathcal{Y}}
\newcommand{\Ypm}{\mathcal{Y}^\pm}

\newcommand{\xt}{x_{\bot}}
\newcommand{\xit}{\xi}
\newcommand{\aH}{\mathbf{H}}
\newcommand{\aW}{\mathbf{W}}

\title{Asymptotic wave-splitting in anisotropic linear acoustics}
\author{B. L. G. Jonsson and M. Norgren\\
Electromagnetic Engineering, \\Royal Institute of Technology,
Stockholm, Sweden}

\begin{document}
\maketitle

\section*{Abstract}

Linear acoustic wave-splitting is an often used tool in describing
sound-wave propagation through earth's subsurface.  Earth's
subsurface is in general anisotropic due to the presence of
water-filled porous rocks. Due to the complexity and the
implicitness of the wave-splitting solutions in anisotropic media,
wave-splitting in seismic experiments is often modeled as isotropic.
With the present paper, we have derived a simple wave-splitting
procedure for an instantaneously reacting anisotropic media that
includes spatial variation in depth, yielding both a traditional
(approximate) and a `true amplitude' wave-field decomposition. One
of the main advantages of the method presented here is that it gives
an {\em
  explicit} asymptotic representation of the linear
acoustic-admittance operator to all orders of smoothness for the
smooth, positive definite anisotropic material parameters considered
here. Once the admittance operator is known we obtain an explicit
asymptotic wave-splitting solution.

\section{Introduction}

The present paper derives an explicit asymptotic representation of
the linear acoustic-admittance operator in an instantaneously
reacting anisotropic media. This solution enables us to obtain an
explicit asymptotic representation of the wave-splitting operators
in such anisotropic media.

Wave-splitting, or wave-field decomposition, is a tool to decompose
the wave-field into `up'- and `down'-going wave field constituents
in configurations with a certain
directionality~\cite{Kristensson+Kruger85,MdeHoop9606,Jonsson+deHoop01,Morro2004},
as \eg~a seismic experiment for probing earth's subsurface
\eg~\cite{MalcolmMdeHoop05}. The wave-splitting procedure results in
two one-way equations for the wave-field constituents.
Wave-splitting has been used to model and analyze wave propagation
in both inverse problems and migration models. The method of
wave-splitting has a long history with a wide area of
applications; an overview of some of the history is given in
\cite{He+Strom+Weston98}. For the isotropic case wave-splitting has
been used extensively to construct fast propagation
methods~\cite{deHoop+Gautesen,Gustafsson2000,LeRousseau2003,Wu2007}.

Recently, there has been an interest in methods that are almost
frequency independent in calculation complexity based on
wave-splitting \cite{Stolk2009,Popovic2009}. Wave-splitting has also
given raise to algorithms to reconstruct material parameters, for
example the generalized Bremmer coupling series approach and the
downward continuation
approach~\cite{MdeHoop9606,MalcolmMdeHoop05,Stolk+deHoop2006} see
also~\cite{Weston93}.  Another application area of wave-splitting is
in the context of boundary conditions and time-reversal
mirrors~\cite{Jonsson2004,Fishman2009}. Wave-splitting methods have
been implemented in several physically different contexts and for a
range of constitutive relations: wave-splitting for wave
equations~\cite{Weston90,WestonJonsson2002}. The electromagnetic
equations are wave-field decomposed both for
isotropic~\cite{LundstedtHe97,Weston2001,Morro2004}, anisotropic
lossless (the spectral theoretical approach)~\cite{Jonsson2008} and
wave-splitting has been extended to the homogeneous lossless
stratified bi-an\-iso\-tro\-pic
case~\cite{Rikte2001,Kristensson2004}. Wave-splitting methods have
been applied to linear-elas\-to\-dy\-na\-mic equations as for
propagation on beams see \eg~\cite{Johansson2006} as well as in the
half-space in homogeneous stratified anisotropic
media~\cite{deHon96} and up-/down symmetric
media~\cite{deHoopSquare}.

One limitation to the present methods of wave-splitting is that it
has been almost exclusively limited to isotropic media see
\eg~\cite{Kristensson+Kruger85,vanStralen97,Cao98,He+Strom+Weston98,Morro2004}.
The underlying reason for this is that the wave-splitting procedures
can be reduced to solving a certain key equation for the linear
acoustic-admittance. This equation is almost trivial to solve in the
case of isotropic media where it reduces to a square root of a
certain elliptic operator. However, in the heterogeneous anisotropic
media, the equation for the acoustic admittance is a non-linear
equation, an operator Riccati equation, and this equation has
largely resisted explicit solutions. It has been shown
in~\cite{deHoopSquare} that wave-splitting methods for the
exceptional case of up-/down symmetric anisotropic materials
resemble the methods for the isotropic case.

The method developed in the present paper solves the above mentioned
equation for the acoustic admittance operator in an inherently
heterogeneous anisotropic and instantaneously reacting media, at the
same time the method yields explicit asymptotic solutions. We expect
that with the here developed methods, wave-splitting in anisotropic
media can provide a starting point for wave-splitting applications
in anisotropic media. We also expect that some of the isotropic
media applications can be extended to anisotropic media.

Before this paper there was essentially one other method for
wave-splitting that has been extended or developed for the
anisotropic case: The spectral theoretical
approach~\cite{deHon96,Jonsson+deHoop01}. This method is a
constructive method, but it rests on a certain spectral projection
of the so called linear acoustic systems matrix, which can be hard
both to evaluate numerically and to extend the method to more
general constitutive relations. The leading order term of the linear
acoustic admittance operator was obtained
by~\cite{Jonsson+deHoop01}, for the case of a symmetric
heterogeneous anisotropic instantaneously reacting media. Higher
order terms did not appear in their work, and seems to be hard to
obtain by the method presented there.

An often occurring approximation when working with wave-splitting is
to ignore a certain lower order `vertical' variation in the so
called decomposition operator, see \eg~\cite{MdeHoop9606}. It is
known that this lower order variation can be accounted for in an
exact decomposition for the isotropic
media~\cite{Weston90,Zhang2003}, the latter introduced the notation
of `true amplitude one-way wave equations'. An alternative approach
is to include the correction term in the sources of the
problem~\cite{MdeHoop9606,Gustafsson2000}. In the present paper we
show that the explicit asymptotic admittance operator can include or
ignore the vertical variation of the decomposition operator with
minimal changes to the solution. We give both the solutions for both
these cases in heterogeneous anisotropic media.

This paper is organized in seven sections and an appendix, where
Section~\ref{sec2} contains the explicit limitations to the material
parameters of the linear acoustic equation considered here, as well
as a reformulation of the linear acoustics into a form more
suitable for wave-splitting. The wave-splitting problem is posed in
Section~\ref{aws}, and reformulated into find a solution of an operator
Ricatti equation for the acoustic admittance. The section ends with
a brief comparison with earlier wave-splitting methods and their
limitations. Before proceeding with the splitting-procedure, we
recall the basics of pseudodifferential operators with parameters in
Section~\ref{sec4}. The key element in this section is, apart from the
introduction of symbols, the asymptotic composition formula of
symbols.

The explicit asymptotic pseudodifferential solution to the operator
Riccati equation is derived in Section~\ref{sec5}. The asymptotic
acoustic-admittance solution is given to all orders of smoothness in
terms of a recursive formula.  The lowest order terms are given
explicitly. The result is compared with the known result for special
cases, and our result is in accordance with and generalizes these
known results. The asymptotic expression for the solution expressed in
the symbol of the acoustic admittance operator is presented in
Section~\ref{sec5b} together with certain smoothness claims.
Section~\ref{sec6} contains conclusions and reflections on our
result. In an appendix we have included a discussion on the underlying
function spaces used in the wave-splitting procedure as well and the
normalization freedom of wave-splitting solutions.

\section{Linear acoustics and the systems matrix}
\label{sec2}

The motion of sound waves propagation through earth's subsurface is
approximated by the linear acoustic
equations~\cite{Landau+Lifshitz66,Brekhovskikh90,Doicu00,ATdeHoop95}.
The linear acoustic equations are a linear system of equations in
time and space which describe the spatial and temporal changes of
the pressure, $p$, and the particle velocity, $v=(v_1,v_2,v_3)$.  In
the present paper we consider the linear acoustic equations under a
time-Laplace transform, yielding the equations of motion in the
form:
\begin{align}\label{1a}
s \kappa(x)  p(x,s) + \partial_{j}v_j(x,s) &= q(x,s), \\
s \rho_{jk}(x)v_k(x,s)+ \partial_{j}p(x,s) &= f_j(x,s),\ j=1,2,3,
\label{1b}
\end{align}
where we have used the summation notation over repeated index
$j,k\in\{1,2,3\}$. This kind of summation notation over repeated
$j,k$ indices is used throughout the paper. Here $x=(x_1,x_2,x_3)$
is a point in space, $\partial_{k}:=\frac{\partial}{\partial x_k}$
and $s\in \mathbb{C}$ is the one-sided Laplace transform coordinate
dual to time. Zero initial conditions have been assumed as to make
$\frac{\partial }{\partial t} \rightarrow s$. Causality of the
motion is taken into account by requiring that the Laplace-domain
quantities are bounded functions of position in all of space for all
$s$ such that $\RE{s}>0$. In addition we assume that
\begin{equation}\label{arg}
 |\arg s| < \frac{\pi}{2}.
\end{equation}
Furthermore, $f_k$ is the volume source density of force, and $q$ is
the volume source density of injection rate.

The scalar compressibility, $\kappa$, and the volume density mass
tensor, $\rho$, are the material coefficients in the equations.
Sound waves in earth's subsurface propagate through, \eg
water-filled porous rocks, hence $\rho$ is assumed to be a filled
$3\times 3$ tensor or equivalently that the medium is assumed to be
anisotropic. Both material parameters are in the linear acoustic
approximation assumed to be instantaneously reacting \ie independent
of $s$, and we assume that they satisfy the inequalities
\begin{equation}\label{ineq}
\begin{split}
0<\kappa_0\leq &\kappa(x)\leq \kappa_1<\infty, \ \forall\ x\in\RR^3 \\
0 < \rho_0|\zeta|^2 \leq
 & \rho_{jk}(x)\zeta_j\zeta_k\leq\rho_1|\zeta|^2<\infty, \ \forall\
\zeta\in \RR^3,\ x\in \RR^3.
\end{split}
\end{equation}
Hence, $\rho$ and its inverse are positive definite,
$\alpha:=\rho^{-1}$, and $\alpha_{33}(x)>0$, which is used
repeatedly throughout the paper. In the upcoming calculations we use
$\alpha$ rather then $\rho$ in \eqref{1b}, that is we
reformulate~\eqref{1b} as
\begin{equation}\label{alpha}
s v_j(x,s)+ \alpha_{jk}\partial_{k}p(x,s) = \alpha_{jk}f_k(x,s), \
j=1,2,3.
\end{equation}
The last assumption on $\kappa$ and $\alpha$ (or $\rho$) is that
they are in $\C{\infty}$ and that all derivatives of $\kappa,\alpha$
are bounded functions. This last assumption simplify the upcoming
microlocal analysis.

We single out depth as the preferred direction of propagation, as
for \eg a seismic experiment on probing a subsurface, and choose a
coordinate system where the $x_3$-axis is parallel with the depth
direction, also called the `vertical'-axis. The wave-splitting
procedure decomposes the wave-field into `up'- and `down'-going
wave-field constituents with respect to the vertical axis. We
reformulate~\eqref{1a} and \eqref{alpha} towards a form where we can
apply this up/down splitting of the field. This rewriting has been
given in \eg~\cite{MdeHoop9606,deHon96,Jonsson+deHoop01}, but since
it is a short derivation we include it here for completeness. The
equations are separated into two parts; the first part consists of
\eqref{alpha} with $j=1,2$ describing how the `horizontal' or
`transverse' velocity components $v_1,v_2$ depend on pressure. The
second part consists of the remaining two equations in~\eqref{1a}
and \eqref{alpha}:
\begin{equation}\label{3}
 \begin{split}
\kappa s p + \partial_{3}v_3 -s^{-1}\partial_\mu (\alpha_{\mu
k}\partial_k p)  & =  q-s^{-1}\partial_\mu (\alpha_{\mu k}f_k),
 \\
s v_3+ \alpha_{3k}\partial_{x_k}p & = \alpha_{3k}f_k,
\end{split}
\end{equation}
where we have used~\eqref{alpha} with $j\in\{1,2\}$ to remove all
occurrences of $v_1$, $v_2$ in \eqref{1a} and that $\RE{s}>0$.

We use the repeated indices $\mu$, $\nu$ to indicate summation over
1, 2, just as we use the repeated indices $j$ ,$k$ to indicate
summation over 1, 2, 3 throughout the paper. Note in particular that
the term $\partial_\mu(\alpha_{\mu k}\partial_k p)$ contain
derivatives with respect to $x_3$. An equivalent matrix formulation
of~\eqref{3} is
\begin{equation}\label{preA}
(\T \partial_3 + \hat\A) F =
\begin{pmatrix}
  q - s^{-1}\partial_\mu (\alpha_{\mu k}f_k) \\
  \alpha_{3k}f_k
\end{pmatrix},
\end{equation}
with $F=(v_3,p)^T$ and
\begin{equation}
\T=\begin{pmatrix} 1 & -\partial_{\mu}(\alpha_{\mu 3}\cdot) \\
0 & \alpha_{33}\end{pmatrix}, \ \ \hat\A =
\begin{pmatrix}
0 & \kappa s - s^{-1}\partial_{\mu}(\alpha_{\mu \nu}\partial_\nu \cdot) \\
s & \alpha_{3\mu}\partial_\mu
\end{pmatrix}.
\end{equation}

The desired form of~\eqref{preA} suitable for wave-splitting is
obtained by freeing $\partial_3$ from $\T$. This is achieved by
observing that the matrix operator $\T$ is invertible since
$\alpha_{33}\neq 0$. We find
\begin{equation}\label{3b}
(\partial_{3} + \A)F = N,
\end{equation}
with $N=\T^{-1}(q-s^{-1}\partial_\mu (\alpha_{\mu
k}f_k),\alpha_{3k}f_k)^T$, and the acoustic systems matrix, $\A$, is
given by
\begin{equation}
  \A = \T^{-1}\hat\A=\begin{pmatrix}
    \partial_{\mu} (\alpha_{\mu 3}\alpha_{33}^{-1}\cdot) &
    s\kappa-s^{-1}\partial_{\mu} (Q_{\mu\nu}\partial_{\nu} \cdot) \\
    s\alpha_{33}^{-1} & \alpha_{33}^{-1}\alpha_{3\mu}\partial_{\mu}
   \end{pmatrix}
= \begin{pmatrix} \A_{11} & \A_{12} \\ \A_{21} &
\A_{22}\end{pmatrix},
\end{equation}
in which $Q_{\mu\nu}=\alpha_{\mu\nu}- \alpha_{\mu 3}
\alpha_{33}^{-1}\alpha_{3\nu}$, $\mu,\nu\in \{1,2\}$. The $2\times
2$ matrix $Q$ is positive definite since the $3\times 3$ matrix
$\alpha$ is positive definite. Infact, $Q$ is the Schur complement
of $\alpha_{33}$ in $\alpha$, see \eg~\cite{Horn+Johnson}. An
explicit proof for a similar case is given below in
Equations~\eqref{QQ} and~\eqref{wQ}. See
also~\cite{Jonsson+deHoop01}.

\section{Acoustic wave-splitting}
\label{aws}

In this section we state the wave-splitting problem for the
equation~\eqref{3b}, and discuss some existing solution methods and
their limitations. We furthermore reformulate the wave-splitting
problem into the problem of solving an operator Riccati equation.

Consider a linear invertible composition operator
$\LL=(\LL^+,\LL^-)$, a $2\times 2$ matrix of operators, where
$\LL^\pm$ is the $2\times 1$ columns of operators, which maps the
up- and down-ward components, $W=(u_+,u_-)^T$, to the wave-field,
by~\footnote{Underlying spaces and the normalization freedom of the
solution is discussed in the appendix.}
\begin{equation}\label{wp1}
  F=\LL W.
\end{equation}
We reformulate~\eqref{3b} in terms of $W$ to find
\begin{equation}\label{newA}
  \LL (\partial_{3}+ \cS) W = N-(1-\eta)(\partial_3 \LL)W,
\end{equation}
where we require $\LL$ and $\cS$ to satisfy
\begin{equation}\label{5}
  \eta(\partial_3 \LL)+ \A\LL = \LL \cS,\ \ \cS=\begin{pmatrix} \cS^+ & 0 \\ 0 & \cS^-
\end{pmatrix}
\end{equation}
for some scalar operators $\cS^\pm$. Here $\eta$ takes the value of
either zero or one. The term $\partial_3\LL$ in~\eqref{newA} and
\eqref{5} is below shown to be a lower order correction term as
compared to the leading order of $\LL \cS$, \eg the operator is
smoothing as compared with $\LL\cS$. Since $\partial_3\LL$ is
smoothing, it is common to ignore it, \eg $\eta=0$, in the
wave-splitting procedure. It is accounted for in this case through a
correction term in the sources. An alternative approach is to consider
the case with $\eta=1$ and adjust $\LL$ with respect to these
perturbations. Below we find solutions $\LL_\eta,\cS_\eta$ of
\eqref{5} for both cases.

There is a family of solutions $\LL_\eta$ which solves~\eqref{5},
these solutions are related to each other by a `normalization' of
$\LL_\eta$. A discussion of this normalization freedom can be found
in the appendix.

To find a solution to \eqref{5} we make the ansatz that the columns
of the composition operator $\LL^\pm_\eta$ can be written as
$\LL^\pm_\eta=(\Ypm_\eta,1)^T$ where $\Ypm_\eta$ is the acoustic
admittance operators, it is also called the acoustic generalization
of the Dirichlet-to-Neumann operator. Inserting this ansatz
into~\eqref{5} and eliminating the occurrences of $\cSpm_\eta$ we
obtain the following operator Riccati equation
\begin{equation}\label{rr}
\Ypm_\eta \A_{21}\Ypm_\eta + \Ypm_\eta \A_{22}-\A_{11}\Ypm_\eta
-\A_{12} -\eta\partial_3\Ypm_\eta= 0, \ \eta\in\{0,1\}
\end{equation}
together with the relation for $\cSpm_\eta$ as
\begin{equation}\label{14}
\cSpm_\eta = \A_{21}\Ypm_\eta + \A_{22}.
\end{equation}
Solving the equation~\eqref{rr} yields a pair of solutions which are
used to find the solution to the wave-splitting problem, \ie once
$\Ypm_\eta$ is determined we find $\LL_\eta,\cS_\eta$ such
that~\eqref{5} is satisfied.

Before proceeding to find an asymptotic solution to~\eqref{rr} let
us shortly discuss the existing methods for solving the
wave-splitting problem, \ie to find $\LL_\eta$, $\cS_\eta$ such
that~\eqref{5} is satisfied for special cases of $\A$.  For $\eta=0$
and the cases when $\A_{11}$ and $\A_{22}$ vanish is the most well
known case and we find solutions to~\eqref{rr} as
\begin{equation}
  \Ypm_{\eta=0,\text{iso}}=\pm \A_{21}^{-1}(\A_{21}\A_{12})^{1/2}
  =\pm s^{-1} \alpha_{33}
  (\alpha_{33}^{-1}(s^2\kappa-\partial_\mu Q_{\mu\nu}\partial_\nu))^{1/2},
\end{equation}
resulting in
\begin{equation}
  \cSpm_{\eta=0,\text{iso}} = \A_{21}\Ypm_{\eta=0,\text{iso}}= \pm
  (\alpha_{33}^{-1} (s^2\kappa -
  \partial_\mu Q_{\mu\nu}\partial_\nu))^{1/2},
\end{equation}
since $\A_{22}=0$. This case includes the isotropic media case,
where $\alpha_{jk}(x)=\rho_{\text{iso}}^{-1}(x)\delta_{jk}$ and the
up/down symmetric case~\cite{deHoopSquare} \ie
$\alpha_{3\mu}=0=\alpha_{\mu 3}$ for $\mu=$1, 2. Note that the
square root of $\A_{21}\A_{12}$ is a square root of an elliptic
operator with parameter, since $Q$ is a positive definite matrix,
$\kappa>0$, $|\arg s|<\pi/2$, and $\RE{s}>0$.

For the case of homogeneous material parameters, \eg $\kappa$ and
$\alpha$ independent of $x$ one can find solutions to the
wave-splitting problem through diagonalization of the $\A$ in
spatial Fourier domain, in this case $\partial_3\LL_\eta =0$.

A spectral theoretical approach to wave-splitting for $\eta=0$
resulting in a splitting-matrix, $\B$, has been considered
by~\cite{deHon96} for stratified layers of homogeneous anisotropic
media with instantaneous reaction, it was extended into
heterogeneous anisotropic instantaneously reacting media
in~\cite{Jonsson+deHoop01}. This rather complex method yields, in a
constructive way, the existence of the wave-splitting. The method
constructs the wave-splitting through essentially a spectral
projector of the acoustic systems matrix. In the linear
acoustics,~\cite{Jonsson+deHoop01} found the leading order term of
the splitting operator, $\B$, and hence of the admittance operator.
A solution, is obtained through the equation
\begin{equation}
\Ypm_0 = \B_{21}^{-1}(\pm I -\B_{22})
\end{equation}
where $\B=\frac{1}{\iu \pi}\int_{\iu\mathbb{R}}
(\A-\lambda)^{-1}\diff \lambda$ is a $2\times 2$ matrix of operators
with elements denoted by $B_{\mu\nu}$ $\mu$, $\nu=$1, 2. The
existence of the splitting matrix was shown under the additional
requirement $\alpha(x)$ is a {\em
  symmetric} positive definite matrix. There are two main restrictions of the above
outlined method.  The first and more serious restriction is that the
splitting matrix is complex to calculate explicitly, in the sense
that it is derived from a spectral projection of a non-self adjoint
operator. The second restriction is the requirement that $\alpha$ is
symmetric limiting the range of material parameters that can be
considered.

Splitting for $\eta=1$ has to some extent been considered in
frequency domain see \eg~\cite{Zhang2003} but it is most extensively
studied in the so called time-domain wave-splitting procedures see
\eg~\cite{He+Strom+Weston98} and references therein. Both approaches
consider isotropic media.

\section{Pseudodifferential preliminaries}
\label{sec4}

Our solution to the wave-splitting problem is based on the theory of
microlocal analysis, in this case the theory of pseudodifferential
operators. Before we enter into the details of finding such a
solution, we need some preliminary notation and basic facts about
pseudodifferential operators with parameters. We follow the
notations of~\cite{Shubin87}.

A point $x=(x_1,x_2,x_3)$ in space will be represented by a
transverse part $\xt$ and a vertical part $x_3$, \eg $x=(\xt,x_3)$.
Let the transverse spatial Fourier transform be given as
\begin{equation}
\hat{u}(\xit,x_3,s)= \int_{\mathbb{R}^2} \lexp{-\iu \xit\cdot
  \xt}u(\xt,x_3,s)\diff \xt,
\end{equation}
where $\xit=(\xi_1,\xi_2)$ is the Fourier dual to $\xt$. Since
$\A_{\mu\nu}$ for all $\mu$ ,$\nu=$1, 2 are differential operators
with smooth coefficients, we find the left-symbol, $a_{12}$, of
$\A_{12}$ by $a_{12}(x,\xit;s)= \lexp{-\iu\xit\cdot
\xt}\A_{12}(x,\partial_1,\partial_2;s)\lexp{\iu\xit\cdot
  \xt}$. Resulting in
\begin{equation}
a_{12}=s\kappa + s^{-1}Q_{\mu\nu}\xi_m\xi_\nu -
s^{-1}(\partial_{\mu}Q_{\mu\nu})\iu \xi_\nu.
\end{equation}
Similarly
\begin{equation}
  a_{11}=\iu \xi_\mu \alpha_{\mu 3}\alpha_{33}^{-1} +
  (\partial_{\mu}
  (\alpha_{\mu 3}\alpha_{33}^{-1})),\
a_{21}=s\alpha_{33}^{-1},\ a_{22}=\iu \xi_\mu
\alpha_{3\mu}\alpha_{33}^{-1}.
\end{equation}
The above symbols are {\em symbols with a parameter}, $s\in\CC$.
Since $s$ corresponds to the time derivative, it is necessary to
consider the parameter $s$ as if it is of the same order as $\xi$.

The symbols are decomposed into order of homogeneity in $s,\xi$. To
illustrate this concept, we separate $a_{12}$ into its
poly-homogeneous components in $s,\xi$: $a_{12}=a_{12;1}+a_{12;0}$,
and similarly for the other components of $a$. The explicit form for
each of the poly-homogeneous terms of $a$ is
\begin{align}
a_{11,1}&=\iu \xi_\mu \alpha_{\mu 3}\alpha_{33}^{-1} &&  a_{11,0}=
\partial_{\mu}
  (\alpha_{\mu 3}\alpha_{33}^{-1}) \\
a_{22,1}&=\iu \xi_\mu \alpha_{3\mu}\alpha_{33}^{-1} &&  a_{21,1}=s\alpha_{33}^{-1} \\
a_{12,1} &= s\kappa + s^{-1}Q_{\mu\nu}\xi_\mu\xi_\nu, &&
a_{12,0}=-s^{-1}\partial_{\mu}Q_{\mu\nu}\iu \xi_\nu.
\end{align}

The composition formula for symbols see \eg \cite{Shubin87} is
formulated as follows: Let $P$ and $Q$ be two (properly supported)
pseudodifferential operators in $\mathbb{R}^2$ with corresponding
symbols $q(\xt,\xi)$, $p(\xt,\xi)$ respectively. The composition
formula yielding the symbol, $r$, corresponding to $R=PQ$, is given
by the asymptotic relation~\cite[I.3.6]{Shubin87},
\cite[I.1.5]{Cordes1995}
\begin{equation}\label{c}
r(\xt,\xi)\sim \sum_{\beta} \frac{1}{\beta!}\left[\partial_\xi^\beta
p(\xt,\xi)\right]\left[(\frac{1}{\iu}\partial_{\xt})^\beta
q(\xt,\xi)\right],
\end{equation}
where $\beta$ is the multi-index $\beta\in \mathbb{N}^2$. Here
$\beta=(\beta_1,\beta_2)$, $\beta! = \beta_1!\beta_2!$, and
$\partial_\xi^\beta =\partial_{\xi_1}^{\beta_1}
\partial_{\xi_2}^{\beta_2}$.

The correspondence between a symbol, $r$, and its operator, $R$,
acting upon some function $u$ is
\begin{equation}
(Ru)(\xt) = \frac{1}{(2\pi)^2}\int_{\RR^2} \lexp{\iu \xt\cdot \xi}
r(\xt,\xi)\hat{u}(\xi)\diff \xi.
\end{equation}
Note that partial differential operators with smooth coefficients
are properly supported, and by a uniformity argument this property
extend to partial differential operators with parameters, it is also
true for each given branch of the square root of elliptical
operators with parameters.

\section{Asymptotics of the acoustic admittance operator}
\label{sec5}

In this section we derive the asymptotic expansion of the admittance
operator $\Y$. This is rather long recursive expressions, where the
next term is expressed in terms of all earlier terms. Once we
derived the general expression we compare the derived result with
existing results for a pair of particular cases. Henceforth we
suppress the index $\eta$ for readability.

Starting from the operator Ricatti equation
\begin{equation}\label{r}
\Y (\A_{21}\Y + \A_{22}) - \A_{11}\Y -\A_{12} -\eta(\partial_3\Y)=
0, \ \eta\in{0,1}
\end{equation}
we are interested in deriving the asymptotic expansion of the symbol
$y$ corresponding to a solution $\Y$ of \eqref{r}. We expect to get
two (separate) solutions $\Y^+$ and $\Y^-$ from~\eqref{r}. The
normalization freedom of $\LL$ imply a normalization freedom of $\Y$
and hence we expect an infinite number of solutions in two equivalence
classes, see Appendix~\ref{app}. However, for our purpose it is
sufficient to find one pair of solutions to~\eqref{r}, \eg $\Ypm$.

To translate~\eqref{r} into an equation for symbols, we repeatedly
use the composition relation~\eqref{c}. Let's denote the symbol
corresponding to $\Y$ by $y$. We start with the middle part
of~\eqref{r}$: \R_1:=-\A_{11}\Y-\A_{12}$, its symbol, $r_1$, has the
asymptotic expansion
\begin{equation}
r_1\sim -\iu \xi_\mu\alpha_{\mu 3}\alpha_{33}^{-1}y
-\partial_\mu(\alpha_{\mu 3}\alpha_{33}^{-1} y)
-s\kappa-s^{-1}Q_{\mu\nu}\xi_\mu\xi_\nu-s^{-1}\partial_\mu Q_{\mu\nu}\iu \xi_\nu
\end{equation}
after simplification. The simple form is due to that $\A_{11}$ is a
first order differential operator.
Similarly we study the first part of~\eqref{r} and denote this term
$\R_2:=\Y(\A_{21}\Y+\A_{22})$, its corresponding symbol, $r_2$, can be
obtained directly from~\eqref{c}. Once again due to the simple form
of $\A_{21}$, we find that
\begin{equation}
r_2\sim \sum_{\beta} \frac{1}{\beta!}(\partial_\xi^\beta y)
(\frac{1}{\iu}\partial_x)^\beta (s\alpha_{33}^{-1}y+\iu\xi_\mu\alpha_{3\mu}\alpha_{33}^{-1})
\end{equation}

Solving the operator Riccati equation on a symbol level is
equivalent to solving the equation
\begin{equation}
 r_1+r_2 - \eta\partial_3 y= 0,
\end{equation}
or equivalently solving
\begin{multline}\label{y}
\sum_{\beta} \frac{1}{\beta!}(\partial_\xi^\beta y)
(\frac{1}{\iu}\partial_x)^\beta (s\alpha_{33}^{-1}y+\iu\xi_\mu\alpha_{3\mu}\alpha_{33}^{-1})
 -\iu \xi_\mu\alpha_{\mu 3}\alpha_{33}^{-1}y
-\partial_\mu(\alpha_{\mu 3}\alpha_{33}^{-1} y)
\\ -s\kappa-s^{-1}Q_{\mu\nu}\xi_\mu\xi_\nu+s^{-1}\partial_\mu Q_{\mu\nu}\iu \xi_\nu
-\eta \partial_3 y \sim 0
\end{multline}
with respect to $y$. We observe that the ansatz $y=y_0 +y_{-1}
+y_{-2} + \ldots$, where each term, $y_{-n}$ is poly-homogeneous in
$\xi,s$ of order $-n$, yields a consistent solution of~\eqref{y}.
With this observation, it is also clear that the term $\partial_3 y$
is a lower order term, see also Section~\ref{sec5b}.

Inserting the expansion of $y$ into~\eqref{y} enable us to solve the
equation in an iterative manner, first we collect all 1:st order
terms:
\begin{equation}\label{y0}
s\alpha_{33}^{-1}y_0^2+\iu\xi_\mu\alpha_{33}^{-1}(\alpha_{3\mu}
-\alpha_{\mu 3})y_0
\sim s\kappa+s^{-1}Q_{\mu\nu}\xi_\mu\xi_\nu
\end{equation}
with solutions
\begin{equation}\label{y0p}
  \boxed{y_0^\pm \sim s^{-1}\big(-\frac{1}{2}\iu \xi_\mu (\alpha_{3\mu}-\alpha_{\mu 3})
  \pm \gamma_1 \big),}
\end{equation}
where
\begin{equation}
\gamma_1:=\alpha_{33}^{1/2}(s^2\kappa+\tilde{Q}_{\mu\nu}\xi_\mu\xi_\nu)^{1/2}
\end{equation}
with
\begin{align}
  \tilde{Q}_{\mu\nu}\xi_\mu\xi_\nu = \big(Q_{\mu\nu}-
  \frac{1}{4}(\alpha_{3\mu}-\alpha_{\mu 3})\alpha_{33}^{-1}
  (\alpha_{3\nu}-\alpha_{\nu 3})\big)\xi_\mu\xi_\nu \\ = \big(\alpha_{\mu\nu}-
  \frac{1}{4}(\alpha_{\mu
    3}+\alpha_{3\mu})\alpha_{33}^{-1}(\alpha_{\nu 3}+
    \alpha_{3\nu})\big)\xi_\mu\xi_\nu.
\end{align}
Note that $\tilde{Q}_{\mu\nu}$ is positive definite, this follows
from the assumption that $\alpha$ is positive definite and the
observation that for any vector $\xi\in \RR^2$ we can find a vector
$\zeta\in\RR^3$ such that~\footnote{Recall that we always sum over
1, 2, 3 for repeated the repeated indices $j$, $k$, and over 1, 2
for the repeated indices $\mu$, $\nu$.}
\begin{equation}\label{QQ}
\tilde{Q}_{\mu\nu}\xi_\mu\xi_\nu = \alpha_{jk}\zeta_j\zeta_k
\end{equation}
and since $\alpha>0$ we find that $\tilde{Q}>0$. The relation
between $\xi$ and $\zeta$ is
\begin{equation}\label{wQ}
\zeta=\big(\xi_1,\xi_2,
  -\frac{1}{2}\alpha_{33}^{-1}(\alpha_{3\mu}+\alpha_{\mu 3})\xi_\mu\big).
\end{equation}
This proof is similar to the proof that if $\alpha>0$ then $Q>0$
given in~\cite{Jonsson+deHoop01}.

The zero order terms in~\eqref{y} are:
\begin{multline}
  \alpha_{33}^{-1}y_{-1}\big(2 s y_0^\pm + \iu \xi_\mu(\alpha_{3\mu}-\alpha_{\mu 3})\big)+
  \sum_{|\beta|=1} \frac{1}{\beta!}(\partial_\xi^\beta y_0^\pm)
  (\frac{1}{\iu}\partial_x)^\beta (s\alpha_{33}^{-1}y_0^\pm+
  \iu\xi_\mu\alpha_{3\mu}\alpha_{33}^{-1}) \\
  -\partial_\mu(\alpha_{\mu 3}\alpha_{33}^{-1} y_0^\pm) - \eta\partial_3 y_0^\pm \sim
  -s^{-1}\partial_\mu Q_{\mu\nu}\iu \xi_\nu.
\end{multline}
Once we observe that $2 s y_0^\pm + \iu
\xi_\mu(\alpha_{3\mu}-\alpha_{\mu 3}) = \pm 2\gamma_1$ and recall
that $\gamma_1$ is the first order symbol of a square root of a
second order elliptic operator with parameters since $|\arg
s|<\pi/2$ and $\RE{s}>0$ and hence invertible~\cite{Shubin87}. We
find that the next order term, $y_{-1}$ has the asymptotic
behavior:
\begin{empheq}[box=\fbox]{multline}\label{yn}
y_{-1}^\pm \sim \pm \frac{\alpha_{33}}{2 \gamma_1} \Big\{
  -s^{-1}\partial_\mu Q_{\mu\nu}\iu \xi_\nu +\partial_\mu(\alpha_{\mu
    3}\alpha_{33}^{-1} y_0^\pm)+\eta\partial_3 y_0^\pm \\- \sum_{|\beta|=1}
  \frac{1}{\beta!}(\partial_\xi^\beta y_0^\pm)
  (\frac{1}{\iu}\partial_x)^\beta (s\alpha_{33}^{-1}y_0^\pm+
  \iu\xi_\mu\alpha_{3\mu}\alpha_{33}^{-1}) \Big\}.
\end{empheq}

We proceed similarly for each poly-homogeneous order in~\eqref{y}
and can thus from the $-n$:th order terms find the
$y_{-n-1}^\pm$ solution for $n>0$. The solutions are expressed in
terms of $y_{-n}^\pm,\ldots y_0^\pm$. Its explicit expression is
\begin{empheq}[box=\fbox]{multline}\label{ynn}
y_{-n-1}^\pm = \pm \frac{\alpha_{33}}{2\gamma_1}\Big\{
\partial_\mu(\alpha_{\mu 3}\alpha_{33}^{-1} y_{-n}^\pm)+\eta\partial_3 y_{-n}^\pm-
s\alpha_{33}^{-1} \mathop{\sum_{j+k=-n-1}}_{-n\leq j,k\leq -1}
y_{j}^\pm y_{k}^\pm \\- \sum_{k=1}^{n+1}\sum_{|\beta|=k}
\frac{1}{\beta!} \mathop{\sum_{j+m=k-n-1}}_{-n\leq j,m\leq
  0}(\partial_\xi^\beta y_j^\pm)(\frac{1}{\iu}\partial_x)^\beta(
  s\alpha_{33}^{-1}y_m^\pm +
\delta_{0m}\iu \xi_\mu\alpha_{3\mu}\alpha_{33}^{-1})\Big\}.
\end{empheq}
With the results in~\eqref{y0p}, \eqref{yn} and~\eqref{ynn} we have
thus obtained an asymptotic representation of $y^\pm$ and hence of
$\Ypm$. This solution solves the operator Riccati equation~\eqref{r}.

Comparing with the case of a symmetric $\alpha=\alpha(x)$ as detailed
in~\cite{Jonsson+deHoop01} we find that their leading order term
agrees with our leading order term for the case when $\eta=0$.
 Furthermore, for the simpler isotropic case the
above result agrees with the leading order term reported
in~\cite{MalcolmMdeHoop05}.

\section{Claim of lower order}
\label{sec5b}

In this section we use the above derived acoustic admittance symbol
to express the respective (matrix) symbols of the wave-splitting
solution. We also show the claim that $\partial_3\LL$ is smoother
than $\LL\cS$.

The above introduce ansatz $\LL^\pm=(\Ypm,I)^T$ does indeed give us
a solution to the splitting problem. Starting from~\eqref{14} we
find that the diagonal terms in $\cS$ have the symbol, $\cs^\pm$
with poly-homogeneous expansion
\begin{multline}\label{gp}
  \cs^\pm = \alpha_{33}^{-1}(s y^\pm + \iu\xi_\mu\alpha_{3\mu})
  \sim \alpha_{33}^{-1}(sy^\pm_0 + \iu \xi_\mu \alpha_{3\mu})
  +\alpha_{33}^{-1}s(y_{-1}^\pm+y_{-2}^\pm+\ldots) \\
  = \cs^\pm_1 + \cs^\pm_0+\cs^\pm_{-1}+\cdots,
\end{multline}
where
\begin{equation}
\cs^\pm_-1=\alpha_{33}^{-1}(sy^\pm_0 + \iu \xi_\mu \alpha_{3\mu}), \
\cs^\pm_{-n} = \alpha_{33}^{-1}s y_{-n-1}^\pm, \ n=0,1,2,\ldots
\end{equation}
Hence $\cs^\pm$ is expressed as a poly-homogeneous sum with
leading order term of poly-homogeneous order one.

Similarly we find that the composition operator $\LL$ with symbol,
$\lu$, of the form:
\begin{equation}\label{lu}
\lu = \begin{pmatrix} y^+ & y^- \\ 1 & 1 \end{pmatrix} \sim
\lu_0+\lu_{-1}+\lu_{-2}+\cdots,
\end{equation}
where
\begin{equation}\label{luz}
\lu_0 = \begin{pmatrix} y^+_0 & y^-_0 \\ 1 & 1 \end{pmatrix},\
\lu_-n = \begin{pmatrix} y^+_{-n} & y^-_{-n} \\ 0 & 0 \end{pmatrix},\ n=0,1,2\ldots
\end{equation}

We claimed in Section~\ref{aws} that the term $\partial_3 \LL$ is of
lower order than $\LL\cS$, we are now in the position to show this
claim. We do this by explicitly giving the symbols of
$\partial_3\LL$ and $\LL\cS$. The symbol of $\partial_3 \LL$ can be
written as
\begin{equation}
\partial_3 \lu  \sim \begin{pmatrix} \partial_3y^+ & \partial_3y^- \\
0 & 0 \end{pmatrix},
\end{equation}
where
\begin{multline}\label{dzy}
\partial_3y^\pm \sim s^{-1}\big(-\frac{1}{2}\iu\xi_\mu\partial_3(\alpha_{3\mu}-\alpha_{\mu 3})
\pm \frac{1}{2\gamma_1}(s^2\partial_3\kappa +\big(\partial_3
\tilde{Q}_{\mu\nu}) \xi_\mu\xi_\nu\big)\Big) \\ + \text{terms of
order -1 and lower}.
\end{multline}
The symbol, $p$, of $P:=\LL\cS$ is
\begin{multline}
p\sim \sum_\beta (\partial_\xi^\beta
\lu)(\frac{1}{\iu}\partial_x)^\beta \cs \sim \lu_0 \cs_1 + \lu_{-1}
\cs_1 + \lu_0 \cs_0 + \sum_{|\beta|=1} (\partial_\xi^\beta
\lu_0)(\frac{1}{\iu}\partial_x)^\beta \cs_1
\\ + \text{terms of order -1 and lower}.
\end{multline}
where $\ell_m$ is the $2\times 2$ matrix see~\eqref{lu} and
$\cs_m=\diag(\cs^+_m,\cs^-_m)$, with $\cs^\pm_m$ for
$m\in{1,0,-1,-2,\cdots}$ as given in ~\eqref{gp}.

Comparing the result for the respective leading orders, of $\partial_3
\ell$ and $p$, we find that  $\partial_3\ell_0$ is a matrix
containing the terms $(\partial_3 y^\pm_0,0)$ while the leading
expression for $p$ is $\lu_0\cs_1$ with terms $(y_0^\pm\cs_1^\pm,\cs_1^\pm)$.
We find the claim to be shown,
$\LL\cS$ has leading order symbol of order one whereas $\partial_3\LL$
has leading order symbol of order zero in the poly-homogeneous expansion of the symbols in $s,\xi$. Hence, $\partial_3 \LL$ is a
smother term as compared with $\LL\cS$ as claimed.

\section{Conclusion}\label{sec6}

The main advantage with the above procedure to derive $\Ypm$ as
compared to the spectral theoretical approach is that it gives an
explicit asymptotic representation of $\Ypm$ through its asymptotic
symbol expansion $y\sim y_0+y_{-1}+\ldots$ as given above. The
method yields, as far as the authors know, the first explicit form
of $y_{-n}$ for $n>0$ in anisotropic media.

The presented method is not limited to material parameters for which
certain spectral properties of $\A$ exist, in our method it suffices
to consider material parameters $\kappa(x),\alpha(x)$ such that
$\alpha_{33}$ and $s^2\kappa
+\partial_{x_\mu}\tilde{Q}_{\mu\nu}\partial_{x_\nu}$ have well
defined inverses and square roots.

The flexibility of the presented approach is shown by the
possibility to include the often neglected term due to the vertical
derivatives of the composition operator with minimal changes to the
derivation procedure. The simplicity of the method indicates that it
might be extended to the case of the electromagnetic equations,
possibly with anisotropic conductivity, as well as to linear
elasticity.

\appendix

\section{Non-uniqueness of wave-splitting solutions}
\label{app}

To formalize our discussion of the normalization freedom of
wave-splitting solutions we introduce the function space
$\aH^n:=(\Sob{n}(\RR^3,\CC), \Sob{n+1}(\RR^3,\CC))^T$, where $\Sob{n}$
is the $n$:th order Sobolev space of square integrable functions,
$n\in\RR$, see \eg \cite{Cordes1995}.

The normalization freedom implicitly appear in the first step in
phrasing the wave-splitting problem:
\begin{equation}\label{5prim}
F=\LL W.
\end{equation}
Here we need to specify to what spaces $F$ and $W$ belongs. Let's
start with the case outlined in the paper. We have implicitly
assumed that the sources $N$ in \eqref{3b} belong to $\aH^{n-1}$,
for some $n\in \RR$ and hence, due to the derivatives in $\A$, we
find that $F\in \aH^{n}$. The obtained $\LL$ with symbol given in
\eqref{lu} maps $\aH^{n+1}$ to $\aH^n$, and the obtained wave-field
constituents satisfy $W\in\aH^{n+1}$. Indeed if we use the notation
$\LL_n$ to indicate that $\LL_n: \aH^n \mapsto \aH^{n-1}$ and
similarly for $\cS$ we can rephrase the requirements on $\LL,\cS$
as: find $\LL,\cS$ such that
\begin{equation}
\eta(\partial_3\LL_{n+1})+ \A \LL_{n+1} = \LL_n\cS_{n+1}, \ \eta\in\{0,1\}
\end{equation}
with $\cS_{n+1}$ diagonal.

To make the normalization freedom apparent we replace \eqref{5prim}
with $F=\tilde{\LL}_{n+1}\tilde{W}$, where
\begin{equation}\label{tilde}
\tilde{\LL}_{m} = \LL_{m} \N_m^{-1}, \ \tilde{W} = \N_{n+1} W,\
m, n\in \RR
\end{equation}
for any invertible operator $\N_m$ from $\aH^m$ to some desired
space $\aW_m$ for some $m\in \RR$.

The natural question arises: what are the requirements on $\N$ so
that $\tilde{\LL}$ and its corresponding $\tilde{\cS}$ is a wave-
splitting solution to \eqref{wp1}-\eqref{5}. These operators $\N$ represents
{\it the normalization freedom} of the solution of the
wave-splitting problem. It is rather straight forward to implicitly
characterize which $\N$ that provides a wave-splitting solution, \eg
we substitute $\LL,\cS$ in \eqref{tilde} into \eqref{newA} and
\eqref{5} to find
\begin{multline}\label{q}
\big(\eta (\partial_3\tilde\LL_{n+1}) + \A \tilde{\LL}_{n+1}\big)\tilde{W} = \\
\tilde{\LL}_n\big[ \N_n\cS_{n+1}\N_{n+1}^{-1} - \eta
(\partial_3\N_{n+1})\N_{n+1}^{-1}\big]\tilde{W} = \tilde{\LL}_n
\tilde{\cS}\tilde{W},
\end{multline}
where we have identified
$\tilde{\cS}_{n+1}:=\N_n\cS_{n+1}\N_{n+1}^{-1}-\eta(\partial_3\N_{n+1})\N_{n+1}^{-1}$.
The requirement on $\N_m$ is that $\tilde{\cS}_{n+1}$ remains diagonal
for diagonal $\cS_{n+1}$.  In addition as an implicit requirement on
$\N$, for $\tilde{W}\in\aW_{n+1}$ we require that
$(\partial_3\N_{n+1})\N_{n+1}^{-1}\tilde{W} \in \aW_n$, as indicated
in the restriction of the $\tilde{\LL}_{n+1}$ term to $\tilde{\LL}_n$
in front of $(\partial_3 \N_{n+1})\N_{n+1}^{-1}$ in \eqref{q}.

The normalization freedom above is rather large, there are several
diagonal (and anti-diagonal) $\N$ which satisfy the above conditions,
\eg $\N=\diag(m,m')$, $m,m'\in \RR$ and
$\N=\diag((\Y^+)^p,(\Y^-)^{p'})$, $p,p'\in \RR$. The
normalization freedom was used in earlier wave-splitting papers, for
the isotropic case \cite{MdeHoop9606} utilize the normalization
freedom to derive wave-splitting solutions with self-adjoint
admittance operator.  Normalization was also discussed in the
anisotropic case \cite{Jonsson+deHoop01} where the spaces $\aH^n$ also
appeared.

One case of some interest is the normalization that switch from the
acoustic-admittance representation to the acoustic-impedance
representation. Here we are interested in a normalization operator
\begin{equation}
\N=\diag((\Y^+) , (\Y^-)),
\end{equation}
and $\tilde{W}\in\aH^n$. Thus $\N_n:\aH^{n}\mapsto\aH^{n-1}$. The
explicit form of the symbol of $\Y^\pm$ ensure the implicit
assumption that $(\partial_3 \N_{n+1})\N_{n+1}^{-1}\tilde{W} \in
\aH^n$ and that $\tilde{\LL}_{n+1}:\aH^n\mapsto\aH^n$. We obtain
\begin{equation}
\tilde{\LL} = \begin{pmatrix} I & I \\ (\Y^+)^{-1} &
(\Y^-)^{-1}\end{pmatrix}
\end{equation}
and
\begin{multline*}
 \tilde{\cS} = \\ \begin{pmatrix} (\Y^+)^{-1}\cS_+(\Y^+)-\eta(\partial_3(\Y^+)^{-1})\Y^+ & 0
 \\ 0 & (\Y^-)^{-1}\cS_-(\Y^-)-\eta(\partial_3(\Y^-)^{-1})\Y^-
\end{pmatrix}.
\end{multline*}


\begin{thebibliography}{10}

\bibitem{Brekhovskikh90}
L.~M. Brekhovskikh and O.~A. Godin.
\newblock {\em Acoustics of Layered Media I}.
\newblock Springer-Verlag, Berlin, 1990.

\bibitem{Cao98}
J.~Cao.
\newblock {\em Applications of 3{D} domain wave splitting to direct and inverse
  scattering}.
\newblock PhD thesis, Royal Institute of Technology, Stockholm, Sweden, 1998.

\bibitem{Cordes1995}
H.~O. Cordes.
\newblock {\em The Technique of Pseudodifferential Operators}.
\newblock London Mathematical Society Lecture Note Series 202. Cambridge
  University Press, Cambridge, 1995.

\bibitem{Doicu00}
A.~Doicu, Y.~Eremin, and T.~Wriedt.
\newblock {\em Acoustic \& Electromagnetic Scattering Analysis}.
\newblock Academic Press, San Diego, 2000.

\bibitem{Weston2001}
I.~Egorov, G.~Kristensson, and V.~H. Weston.
\newblock Transient electromagnetic wave propagation in laterally
  discontinuous, dispersive media.
\newblock {\em Wave Motion}, 33(1):67--77, 2001.

\bibitem{Fishman2009}
L.~Fishman, B.~L.~G. Jonsson, and M.~V. de~Hoop.
\newblock Time reversal mirrors and cross correlation functions in acoustic
  wave propagation, mathematical modeling of wave phenomena.
\newblock In {\em 3rd Conference on Mathematical Modeling on Wave Phenomena},
  volume 1106 of {\em AIP Conf. Proceeding}, pages 183--202, 2009.

\bibitem{Gustafsson2000}
M.~Gustafsson.
\newblock The {B}remmer series for a multi-dimensional acoustic scattering
  problem.
\newblock {\em J. Phys. A: Math. Gen.}, 33(9-10):1921--32, 2000.

\bibitem{He+Strom+Weston98}
S.~He et~al.
\newblock {\em Time domain wave-splitting and inverse problems}.
\newblock Oxford University Press, Oxford, 1998.

\bibitem{deHon96}
B.~P. {\SortNoop{Hon}{de Hon}}.
\newblock {\em Transient Cross-Borehole Elastodynamic Signal Transfer Through a
  Horizontally Stratified Anisotropic Formation}.
\newblock PhD thesis, Technische Universiteit, Delft, Holland, 1996.

\bibitem{ATdeHoop95}
A.~T. {\SortNoop{Hoop A}{de Hoop}}.
\newblock {\em Handbook of Radiation and Scattering of Waves}.
\newblock Academic Press, Kent, 1995.

\bibitem{MdeHoop9606}
M.~V. {\SortNoop{Hoop AM}{de Hoop}}.
\newblock Generalization of the {B}remmer coupling series.
\newblock {\em J. Math. Phys.}, 37(7):3246--3282, 1996.

\bibitem{deHoop+Gautesen}
M.~V. {\SortNoop{Hoop Gaut}{de Hoop}} and A.~K. Gautesen.
\newblock Uniform asymptotic expansion of the generalized {B}remmer series.
\newblock {\em SIAM J. Appl. Math.}, 60(4):1302--1329, 2000.

\bibitem{deHoopSquare}
M.~V. {\SortNoop{Hoop Hoop}{de Hoop}} and A.~T. de~Hoop.
\newblock Elastic wave up/down decomposition in inhomogeneous and anisotropic
  media: An operator approach and its approximations.
\newblock {\em Wave Motion}, 20:57--82, 1994.

\bibitem{Horn+Johnson}
R.~A. Horn and C.~A. Johnson.
\newblock {\em Matrix Analysis}.
\newblock Cambridge University Press, Cambridge, 1985.

\bibitem{Johansson2006}
M.~Johansson, P.~D. Folkow, and P.~Olsson.
\newblock Dispersion free wave-splitting for structural elements.
\newblock {\em Comput. Struct.}, 84(7):514--527, 2006.

\bibitem{Jonsson2008}
B.~L.~G. Jonsson.
\newblock Wave splitting of {M}axwell's equations with anisotropic
  heterogeneous constitutive relations.
\newblock {\em Inverse Probl. Imag.}, 3(3):405--452, 2009, math-ph/0809.0789.

\bibitem{Jonsson2004}
B.~L.~G. Jonsson, M.~Gustafsson, V.H. Weston, and M.~V. de~Hoop.
\newblock Retrofocusing of acoustic wave fields by iterated time reversal.
\newblock {\em SIAM J. of Appl. Math.}, 64(6):1954--86, 2004.

\bibitem{Jonsson+deHoop01}
B.~L.~G. Jonsson and M.~V. \SortNoop{Hoop}{de Hoop}.
\newblock Wave field decomposition in anisotropic fluids: A spectral theory
  approach.
\newblock {\em Acta Appl. Math.}, 62(2):117--171, 2001.

\bibitem{Kristensson+Kruger85}
G.~Kristensson and R.J. Kruger.
\newblock Direct and inverse scattering in the time domain for a dissipative
  wave equation. {I}. {S}cattering operators.
\newblock {\em J. Math. Phys.}, 27(6):1667--1682, 1986.

\bibitem{Kristensson2004}
G.~Kristensson, S.~Poulsen, and S.~Rikte.
\newblock Propagators and scattering of electromagnetic waves in planar
  bianisotropic slabs --- an application to frequency selective structures.
\newblock {\em PIER}, 48:1–25, 2004.

\bibitem{Landau+Lifshitz66}
L.~D. Landau and E.~M. Lifshitz.
\newblock {\em Fluid Mechanics}.
\newblock Pergamon Press, Oxford, 3:rd edition, 1966.

\bibitem{LundstedtHe97}
J.~Lundstedt and S.~He.
\newblock Time domain direct and inverse problems for a nonuniform {LCRG} line
  with internal sources.
\newblock {\em IEEE T. Electromagn. C}, 39(2):79--88, 1997.

\bibitem{MalcolmMdeHoop05}
A.~E. Malcolm and M.~V. de~Hoop.
\newblock A method for inverse scattering based on the generalized bremmer
  coupling series.
\newblock {\em Inverse Problems}, 21:1137--1167, 2005.

\bibitem{Morro2004}
A.~Morro.
\newblock One-way propagation in electromagnetic materials.
\newblock {\em Math. Comput. Model.}, 39(11-12):1221--29, 2004.

\bibitem{Popovic2009}
J.~Popovic.
\newblock A fast method for solving the {H}elmholtz equation based on wave
  splitting.
\newblock Licentiate thesis, Royal Institute of Technology, Stockholm,
  2009-06-12.

\bibitem{Rikte2001}
S.~Rikte, G.~Kristensson, and M.~Andersson.
\newblock Propagation in bianisotropic media --- reflection and transmission.
\newblock {\em IEE Proc.-Microw. Antennas Propag.}, 148(1):29--36, 2001.

\bibitem{LeRousseau2003}
J.~H. {\SortNoop{Rousseau Hoop}{Le Rousseau}} and M.~V. de~Hoop.
\newblock Generalized-screen approximation and algorithm for the scattering of
  elastic waves.
\newblock {\em Q. J. Mech. Appl. Math.}, 56(1):1--33, 2003.

\bibitem{Shubin87}
M.~A. Shubin.
\newblock {\em Pseudodifferential Operators and Spectral Theory}.
\newblock Springer-Verlag, Berlin, 1987.

\bibitem{Stolk2009}
C.~C. Stolk.
\newblock A fast method for linear waves based on geometrical optics.
\newblock {\em SIAM J. Numer. Anal}, 47(2):1108--1194, 2009.

\bibitem{Stolk+deHoop2006}
C.~C. Stolk and M.~V. de~Hoop.
\newblock Seismic inverse scattering in the downward continuation approach.
\newblock {\em Wave Motion}, 43:579--598, 2006.

\bibitem{vanStralen97}
M.~J.~N. {\SortNoop{Stralen A}{van Stralen}}.
\newblock {\em Directional decomposition of electromagnetic and acoustic
  wave-fields}.
\newblock PhD thesis, Technische Universiteit, Delft, Holland, 1997.

\bibitem{Weston90}
V.~H. Weston.
\newblock Invariant imbedding for the wave equation in three dimensions and the
  applications to the direct and inverse problems.
\newblock {\em Inverse Problems}, 6(6):1075--1105, 1990.

\bibitem{Weston93}
V.~H. Weston.
\newblock Time-domain wave splitting of {M}axwell's equations.
\newblock {\em J. Math. Phys.}, 34(4):1370--1392, 1993.

\bibitem{WestonJonsson2002}
V.~H. Weston and B.~L.~G. Jonsson.
\newblock Wave front layer stripping approach to inverse scattering for the
  wave equation.
\newblock {\em J. Math. Phys.}, 43(10):5045--59, 2002.

\bibitem{Wu2007}
R.-S. Wu, X.-B. Xie, and X.-Y. Wu.
\newblock One-way and one-return approximations ({De Wolf} approximation) for
  fast elastic wave modeling in complex media.
\newblock {\em Adv. Geophys.}, 48:265--322, 2007.

\bibitem{Zhang2003}
Y.~Zhang, G.~Zhang, and N.~Bleistein.
\newblock True amplitude wave equation migration arising from true amplitude
  one-way wave equations.
\newblock {\em Inverse Problems}, 19(5):1113--1138, 2003.

\end{thebibliography}

\newcommand{\SortNoop}[2]{#2}

\end{document}